\documentclass[]{spie}  

\usepackage{amsmath,amsfonts,amssymb}
\usepackage{graphicx}
\usepackage[colorlinks=true, allcolors=blue]{hyperref}
\usepackage{float}
\usepackage{wrapfig}

\title{The Baryon Mapping Experiment (BMX), a 21cm intensity mapping pathfinder}

\author[a]{Paul O'Connor}
\author[b]{An\v{z}e Slosar}
\author[c]{Maile Harris}
\author[a]{Justine Haupt}
\author[a]{John Kuczewski}
\author[c]{Emily Kuhn}
\author[c]{Laura Newburgh}
\author[c]{Annie Polish}
\author[b]{Benjamin Saliwanchik}
\author[a]{Christopher Sheehy}
\author[a]{Paul Stankus}
\author [d]{Gregory Troiani}
\author[c]{Will Tyndall}

\affil[a]{Instrumentation Division, Brookhaven National Laboratory, Upton NY 11973}
\affil[b]{Physics Department, Brookhaven National Laboratory, Upton NY 11973}
\affil[c]{Department of Physics, Yale University, New Haven, CT 06511}
\affil[d]{University of Missouri at Kansas City, Kansas City, MO 64108}

\authorinfo{Further author information: send correspondence to:\\PO: E-mail: poc@bnl.gov, Telephone: 1 631 344 7577\\  }

\begin{document} 
\maketitle

\begin{abstract}
The Baryon Mapping eXperiment (BMX) is an interferometric array designed as a pathfinder for a future post-reionization 21~cm intensity mapping survey. It consists of four 4-meter parabolic reflectors each having offset pyramidal horn feed, quad-ridge orthomode transducer, temperature-stabilized RF amplification and filtering, and pulsed noise injection diode. An undersampling readout scheme uses 8-bit digitizers running at 1.1~Gsamples/sec to provide access to signals from 1.1 - 1.55~GHz (third Nyquist zone), corresponding to HI emission from sources at redshift $0 < z < 0.3$. An FX correlator is implemented in GPU and generates 28~GB/day of time-ordered visibility data. About 7,000 hours of data were collected from Jan. 2019 - May 2020, and we will present results on system performance including sensitivity, beam mapping studies, observations of bright celestial targets, and system electronics upgrades. BMX is a pathfinder for the proposed PUMA intensity mapping survey in the 2030s.
\end{abstract}


\section{INTRODUCTION \& SCIENTIFIC MOTIVATION}
\label{sec:intro}  
To ``Understand Dark Energy and Inflation'' was identified as one of the DOE's five priority science missions in the 2014 report of the High Energy Physics ``P5'' prioritization panel. Since then, optical instruments (such as the Dark Energy Survey \cite{1708.01530}, Dark Energy Spectroscopic Instrument \cite{1611.00036}, Vera Rubin Observatory's LSST \cite{0912.0201} and others) have begun observing and are expected to lead to major advances in the field. More recently, 21\,cm intensity mapping has been gaining attention as a promising new method, complementary to optical and CMB surveys, for studying the nature and evolution of Dark Energy and for investigating signatures of inflation. The intensity mapping technique uses neutral hydrogen as a tracer to map the large scale structure of the Universe by employing  precisely calibrated radio interferometers operating between $\sim$200~MHz and $\sim$1.4~GHz to measure the redshifted emission of the neutral hydrogen spin-flip transition. Source redshifts in the range 0 $\le z \le 6$ are thus unambiguously resolved, since there are no interfering lines. These instruments are very different from traditional radio telescopes, which have much higher angular resolution, smaller collecting area, and less precise calibration requirements than what is needed for 21\,cm IM. Moreover, in intensity mapping experiments, individual sources are not resolved -- instead the number density of sources across space and time is traced by the cumulative emission in the rest-frame 21\,cm transition.  A significant advantage of 21\,cm experiments is that they will benefit from the dramatic cost/performance improvements in digital RF and high-performance computing technologies, driven by the explosive growth of commercial wireless telecommunications technology and machine learning. 

In 2010 a 4$\sigma$ detection of the cosmological large scale structure signal in 21\,cm IM was reported, in an experiment at the Green Bank telescope (GBT) \cite{2010Natur.466..463C}. The authors cross-correlated the radio intensity map with the DEEP2 optical galaxy redshift survey to robustly reject residual systematics from either survey. The 21\,cm signal in auto-correlation from current-generation observations remains significantly contaminated by foregrounds, but the obstacles to detection are purely technical in nature. Current efforts include CHIME \cite{1406.2288}, HIRAX \cite{1607.02059}, and BINGO \cite{1803.01644} among others. We expect results from these experiments to become competitive over the next few years. 

Work towards a large next generation US funded Stage {\sc II} experiment was started a few year ago within the DOE Dark Energy Cosmic Visions group  \cite{1810.09572}. Several interesting science objectives have been identified, among them measuring the expansion history and growth of structure across cosmic ages, as well as probing inflationary physics. Owing to their great sensitivity, these instruments are naturally very competitive also in non-DOE areas of interest, such as the search for Fast Radio Bursts and monitoring pulsars. A concrete implementation of this concept has been proposed in the Packed Ultra-wideband Mapping Array (PUMA) \cite{1907.12559,2002.05072}. 

It is now clear that major technical obstacles remain to be solved before 21\,cm intensity mapping is ready for prime time. We have therefore embarked on designing and building a small prototype instrument situated at the Brookhaven National Laboratory grounds as a testbed for these future technologies. The main motivation was to have a real, operating system, which is conveniently located for agile testing and debugging and contains many of the same features as a Stage {\sc II} experiment: large bandwidth,  low-cost construction, closed-packing of interferometric elements, and transit array observation modes. The main purpose of this instrument is to act as a small-scale test-bed for future technologies allowing fast turn-around times. In particular, we wanted to be able to investigate:
\begin{itemize}
    \item The possibility of operating a radio telescope in an RFI unfriendly environment to learn about possible RFI mitigation techniques
    \item Possible high precision measurements of the beam responses using sources suspended on unmanned aerial vehicles (UAV) such as hexacopter drones and small fixed-wing planes
    \item The overall system stability and various approached towards complex gain calibration
    \item The usability of navigational satellite signals for clock synchronization and possibly relative beam calibration.
\end{itemize}

Based on these requirements we have designed BMX, a 4-dish interferometer testbed. The instrument is described in the next section with some results in Section~\ref{sec:applications}.

%
\section{DESCRIPTION OF THE SYSTEM}
\label{sec:description}  
\begin{figure} [H]
   \begin{center}
   \begin{tabular}{c} 
    \includegraphics[width=0.85 \textwidth]{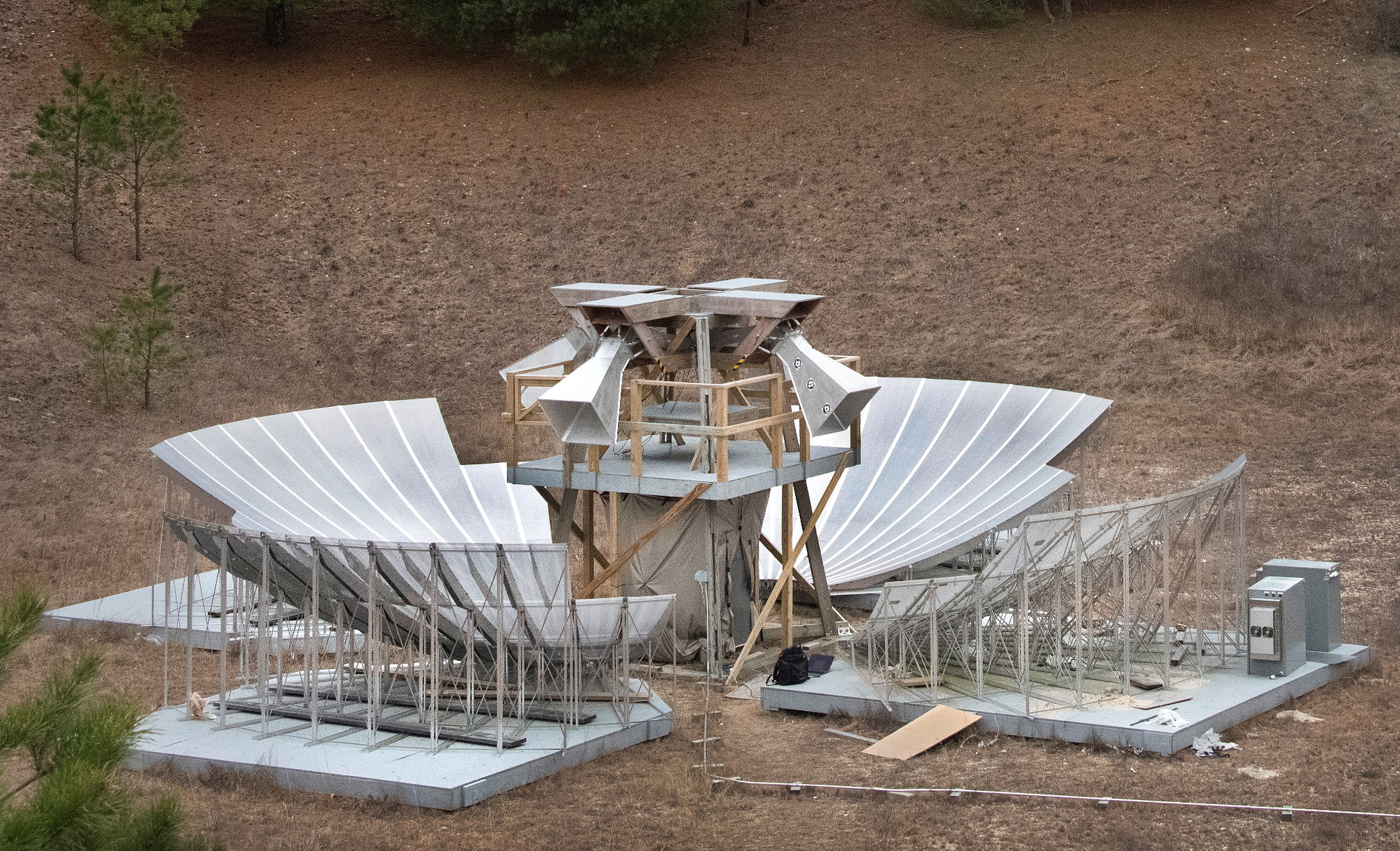}
   \end{tabular}
   \end{center}
   \caption[] 
   { \label{fig:BMX_photo} 
View of BMX installation from the Northeast.}
   \end{figure}

\begin{figure} [h]
   \begin{center}
   \begin{tabular}{c} 
    \includegraphics[width=0.85 \textwidth]{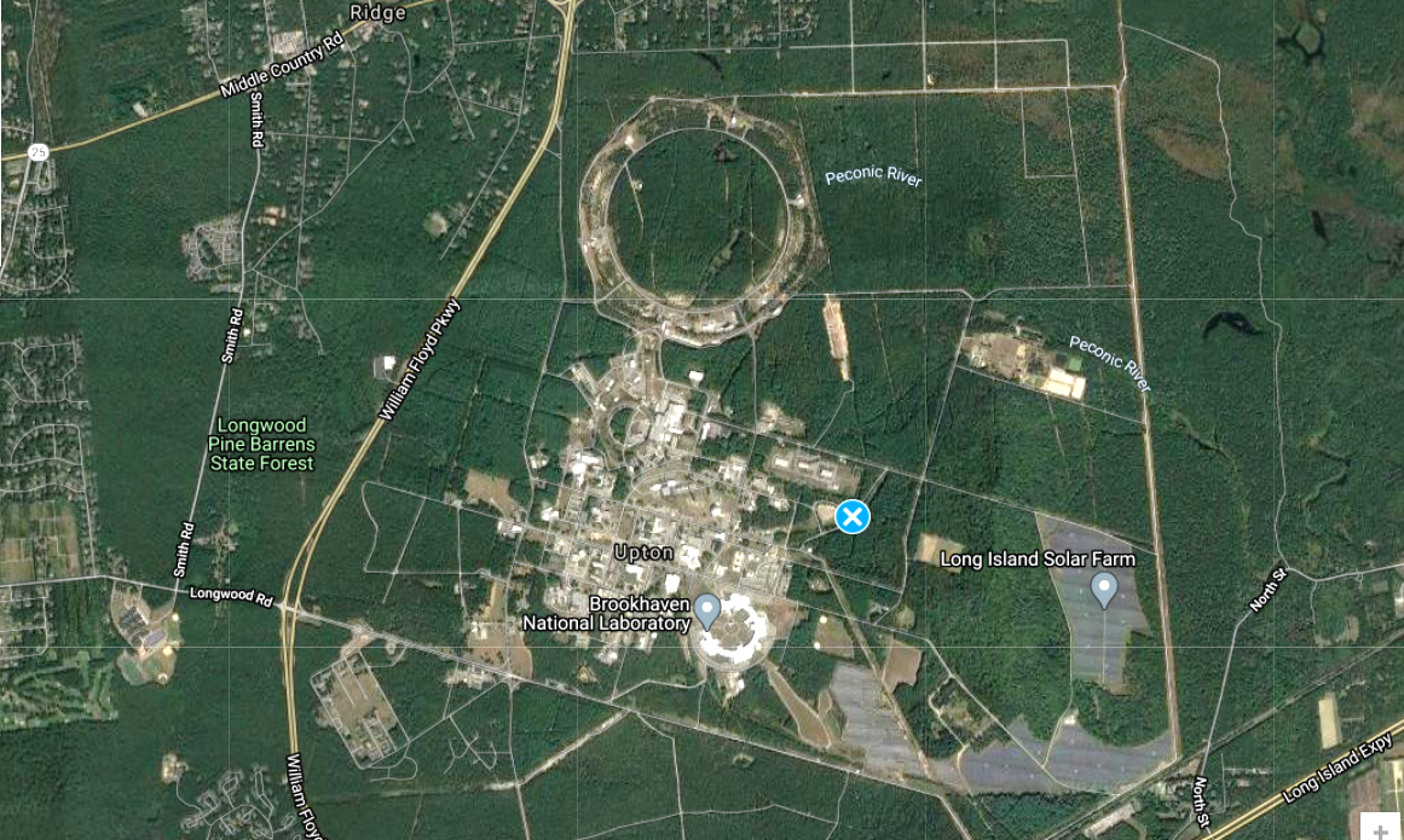}
   \end{tabular}
   \end{center}
   \caption[] 
   { \label{fig:BMX_site} 
BMX site on the BNL campus.}
   \end{figure}

BMX is configured as an array of four, 4-m off-axis parabolic reflectors, each with a pyramidal feedhorn, quad-ridge orthomode transducer (OMT), and dual-polarization receiver operating in the 1.1 - 1.55~GHz band (Fig. \ref{fig:BMX_photo}). The array is located at (40.869929N, 72.866062W) on the Brookhaven Lab campus (Fig. \ref{fig:BMX_site}); the site was chosen after it  was found to have an acceptable RFI environment while providing convenient access for maintenance and upgrades and having AC power and data connections already installed.  The four dishes are built as a square array along cardinal directions with ~8.8 m N-S and E-W baselines. Digitizer and correlator electronics (described in Section \ref{sec:DAQ}) are housed in a shielded, weatherproof enclosure at the base of the central tower and Ethernet-over-power is used to connect to the campus network. BMX was originally built as a single-dish spectrometer in 2017, and upgraded to the interferometer configuration in early 2019. Since then we have logged about 290 days of visibility observations of the $\sim$4 degree wide strip at {$\delta$}=+40.9$^{\circ}$ which passes over the BMX zenith.

%
   
%
%
\subsection{Telescope Geometry, Reflectors and OMT}
\begin{figure}[H]
\begin{center}
\includegraphics[width=0.50 \textwidth]{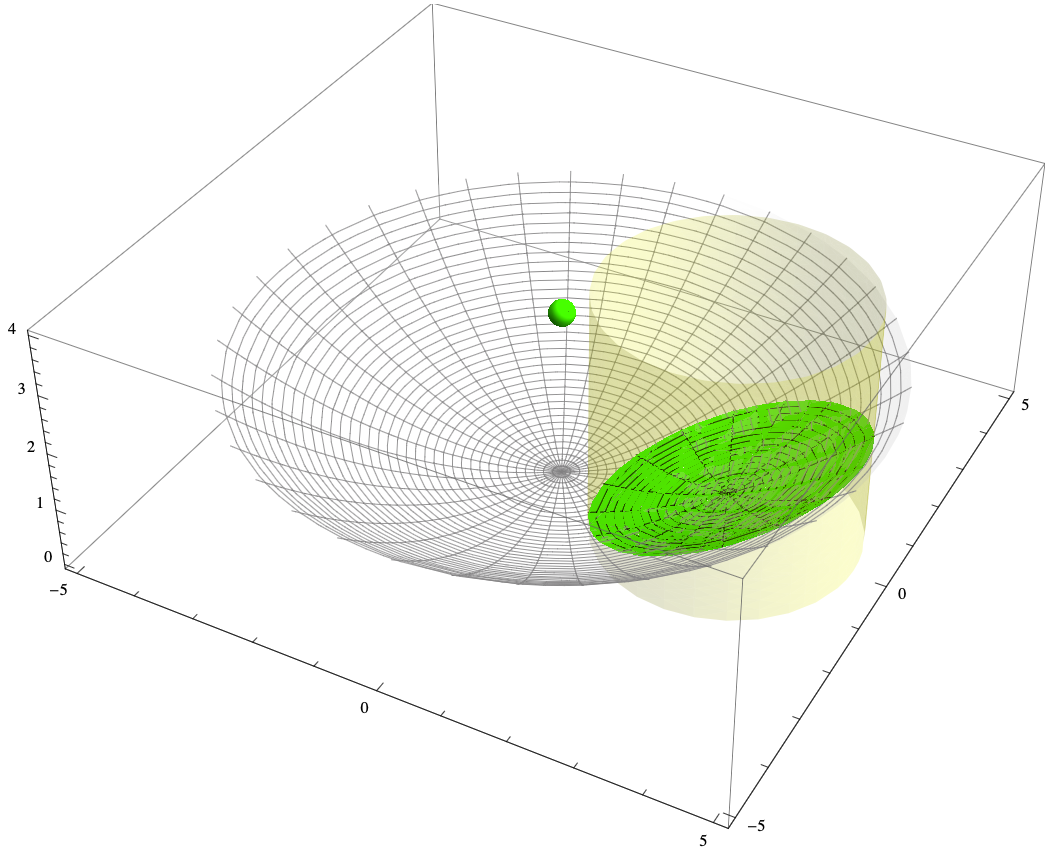}
\includegraphics[width=0.40 \textwidth]{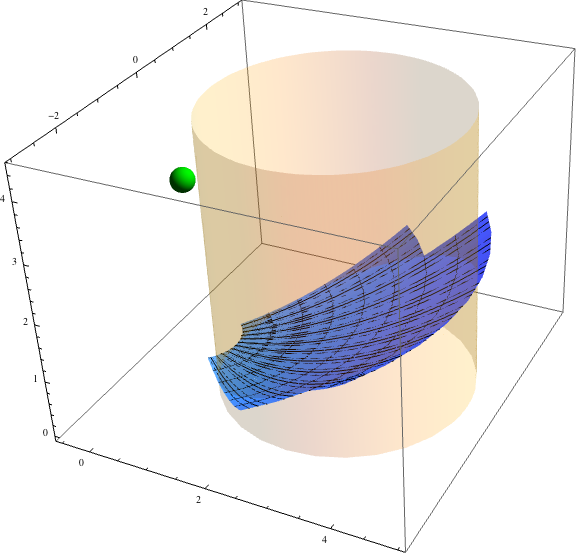}
\end{center}
\caption[]{Left: The nominal 4~m wide, zenith-pointing beam is shown as a cylinder.  The green-shaded  bowl shows the beam's intersection with a parent paraboloid, and so represents the ideal off-axis reflector.  The green dot marks the paraboloid focus at 2.9~m above its vertex. Right: The ideal reflector is approximated as an array of ``petals'', e.g. radial wedge segments of the parent, shown here in blue.  The resulting reflector circumscribes the 4~m vertical beam and brings it to the focus outside the beam path.  Units are in meters.
\label{fig:dish_parent} 
}
\end{figure}

The basic goal that set the reflector and OMT design was (i)~to capture a 4~m wide circular beam directed at the zenith, and (ii)~to do so with the beam being completely unobstructed.  Accordingly an off-axis paraboloid arrangement was chosen, as laid out in Figure~\ref{fig:dish_parent}. For each antenna, the main reflector approximates an off-center cut-out from a parent paraboloid.  The 3-D focal point of this off-axis dish is then the same as that of the parent and lies outside the beam cylinder.  An overall optimization led to the choice of a 2.9~m focal length, as shown in the figure.

%

\begin{figure}[H]
\begin{center}
\includegraphics[width=0.50\textwidth]{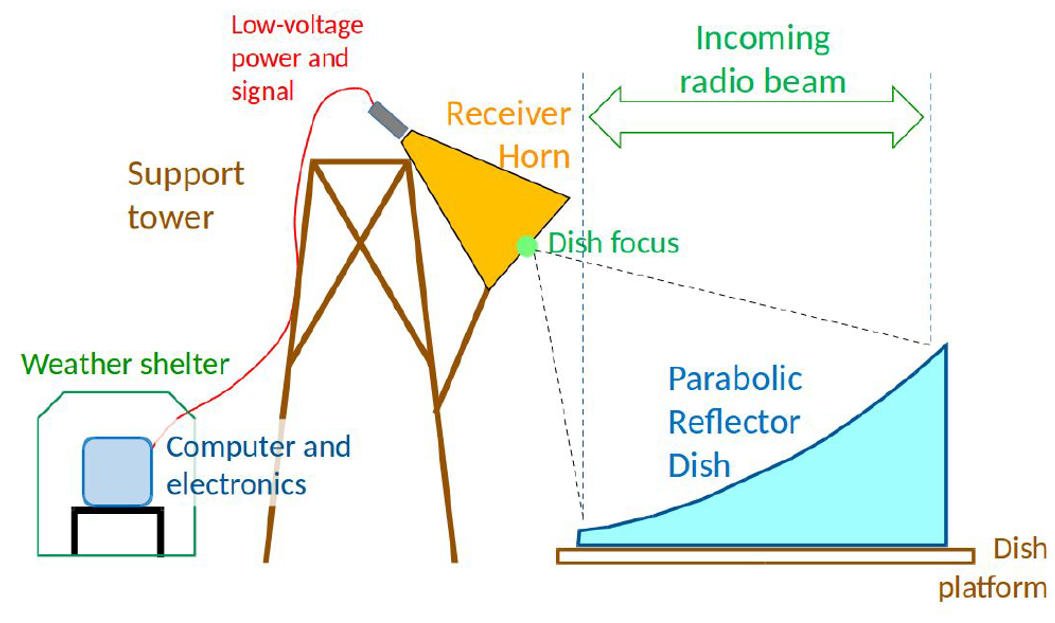}
\includegraphics[width=0.35\textwidth]{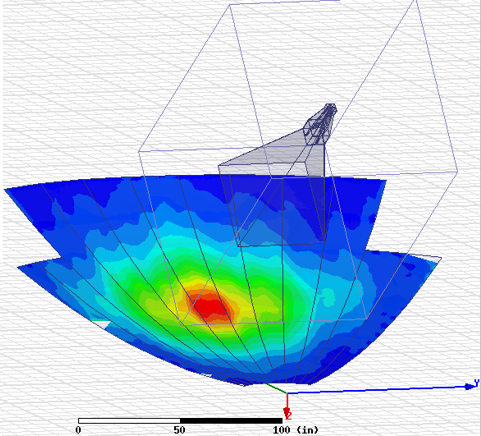}
\end{center}
\caption[]{Left: Sketch of the elements for one antenna.  The zenith-pointing beam is focused by the off-axis parabolic reflector into the pyramidal horn containing the parent paraboloid's focus (see Figure~\ref{fig:dish_parent}). The horn then feeds into the ortho-mode-transducer (OMT) which couples the received power out via coaxial cable.  The dimensions are chosen so that the beam is completely un-obscured by the horn and support structure.  Right: The reflector, horn and OMT are shown in an electromagnetic simulation of the beam system.  The colored ``heat pattern'' shows the illumination of the horn+OMT onto the reflector, showing it is well contained and in fact somewhat under-illuminated.
\label{fig:dish_OMT_tower} 
}
\end{figure}

The reflector was built up out of a series of flat sheets, here termed ``petals'', visible in the right sides of Figures~\ref{fig:dish_parent} and~\ref{fig:dish_OMT_tower}.  Each petal occupies a narrow 15$^o$-wide radial wedge, and is curved in only one direction to avoid the cost and complexity of a compound curved surface. Photogrammetry measurements of the first completed reflector showed $\pm2$mm deviation from a parabolic surface.

%


%
%
\subsection{Front-end amplification}
\begin{figure} [H]
   \begin{center}
   \begin{tabular}{c} 
   \includegraphics[width=\textwidth]{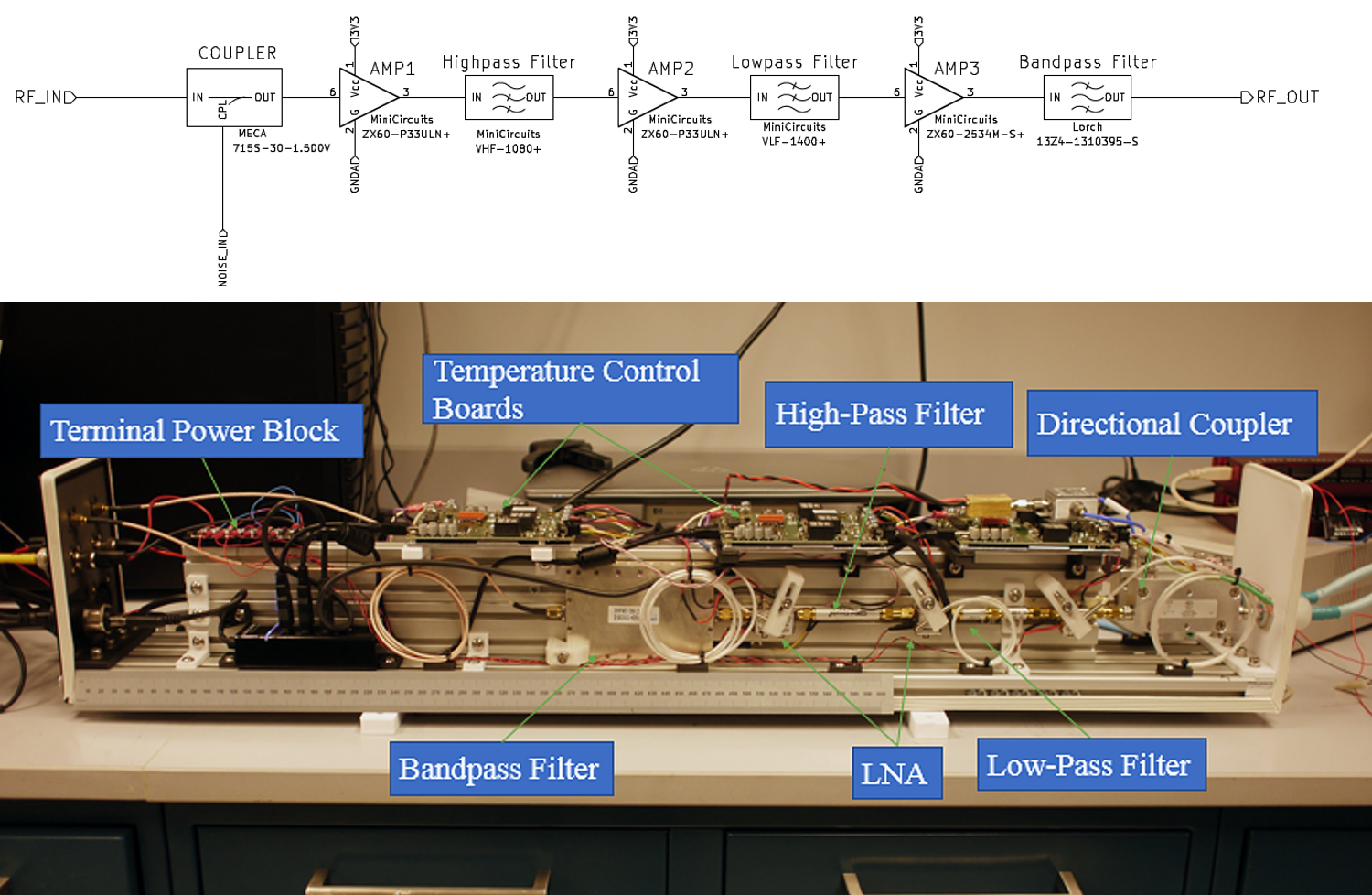}
   \end{tabular}
   \end{center}
   \caption[] 
   { \label{fig:FEE_box} 
Dual-polarization front-end electronics: schematic of one channel (upper), photo with shielded enclosure removed (lower).}
   \end{figure}

The front end electronics boxes are mounted directly behind the horn/OMT assemblies and receive the RF output of each OMT via low-loss coax. Inside the front end box identical amplification and filtering circuits for each polarization are mounted to each side of a rail which provides mechanical support. Peltier elements are sandwiched between the amplifier packages and the rail to allow closed-loop thermal control (not yet activated). A -30dB coupler in front of the first LNA allows broadband noise from a calibrated source to be periodically injected for gain determination. A cavity filter with stopband attenuation of -70dB defines the 1100 - 1550MHz bandpass. The overall chain provides 74dB of gain at the lower band edge, decreasing monotonically by about 7dB across the band. Noise temperature for the signal chain (including directional coupler) measured in the laboratory is 36K at 1100MHz, increasing to 40K at 1500MHz. The OMT and 0.3m low-loss coax are expected to add another 17K. System temperature is measured to be around 70K including losses in the horn and beam spillover to the ground. Power consumption is about 1W per channel, for a total front end power of 8W for all four receivers. A schematic diagram of one channel and a photo of one of the amp boxes is shown in Fig. \ref{fig:FEE_box}. 

\subsection{Data acquisition and Daily Reduction}
\label{sec:DAQ}
Data is acquired by two commercial PCs, each taking care of one polarization mode. Looking down at the telescope from the above, there are two polarization modes, corresponding to the electic field oscillating N-S and E-W. Note that the N-S mode maps on the vertical horn polarization on the N and S dish and horizontal polarization for the E and W dishes. A mirror arrangement holds for the E-W polarization. 

Each PC is equipped with two Spectrum  M4-2223-x8 digitizers. Each digitizer cards supports streaming two channels at 1.1GS/s rates with 8-bit precision. While Spectrum provides 4-channel cards with the same sampling rate, the bus connection at PCI Express x8 Gen 2 would not allow a continuous streaming without loss of data.  The cards are designed to sample at 1.25GS/s and we are undercloking them using an external clock source. Despite being clocked outside their nominal range we have founds the digitizers to be extremely stable. The data acquisition software streams raw samples to the main computer memory and then to the GPU for further processing. 

The 1.1GS/s data rate gives a 500MHz Nyquist frequency and our bandpass defining filter at 1.1-1.55GHz means that we sample the signal in the third Nyquist zone. Data are processed in GeForce GTX 1050 Ti cards using custom CUDA code in chunks of $2^{25}$ samples, corresponding to approximately 30 ms of data, although this is a configurable parameter. We have implemented an asynchronous data transfer and compute on the GPU using CUDA streaming functionality. That is, while one data buffer is being processed, the other data buffer is being transferred onto the graphics cards. The steps in the processing are as follows:\
\begin{itemize}
\item Transfer 4-channels worth of data from main memory to GPU ($\sim 16$ ms)
\item Convert data from 8 bit integer ADC samples to 32 bit floats ($\sim 2$ms)
\item FFT all 4 buffers ($\sim 14$ms)
\item Calculate all 4 auto and 6 complex cross-correlations ($\sim 6$ms)
\item Transfer reduced data back (negligible time)
\end{itemize}
The data is averaged in so-called ``cuts''. For example, during typical operation we average in $2^{13}=8192$ FFT measurements across the full band, giving 2048 spectral channels and additionally a finer grid of $2^9=512$ FFT measurements for fine spectral resolution zoom in around the galactic 21\,cm line.

Because we employ long FFT buffers, the frequency aliasing after averaging is minimal and in fact considerably better than in typical polyphase-filterbank based approaches (which are, however, computationally much cheaper).

We further average 32 individual 30ms measurements into a sample that is saved to disk. This results in approximately 1GB of data produced every hour.
Before each sample is saved to disk we also perform as simple RFI reduction algorithm as follows:
\begin{itemize}
    \item For each spectral bin, we calculate the median signal. 
    \item All samples below the median are used to estimate the variance of the signal around the median.
    \item Any sample that is greater than 4 standard deviations above the median is removed from the average and stored separately an ``RFI rejects'' file.
\end{itemize}
This RFI cleaning approach is motivated by the fact that RFI always adds power rather than removing it. Therefore, for short duration RFI, samples at or below the median level should not be affected. This removes approximately 1\% of the data. The ``RFI rejects'' file is structured to be space efficient (it stores indices and rejected means) allowing one to recover the signal prior to RFI rejection. 

The data are transferred to a local cluster every day and processed overnight using cron jobs. The daily processing involves collating data into a useful format and calculating auxiliary information such as GNSS satellite transits. 

\section{Applications}
\label{sec:applications}
\subsection{Milky Way Analysis}

\begin{figure} [H]
   \begin{center}
  \begin{tabular}{c} 
   \includegraphics[width=0.7\textwidth]{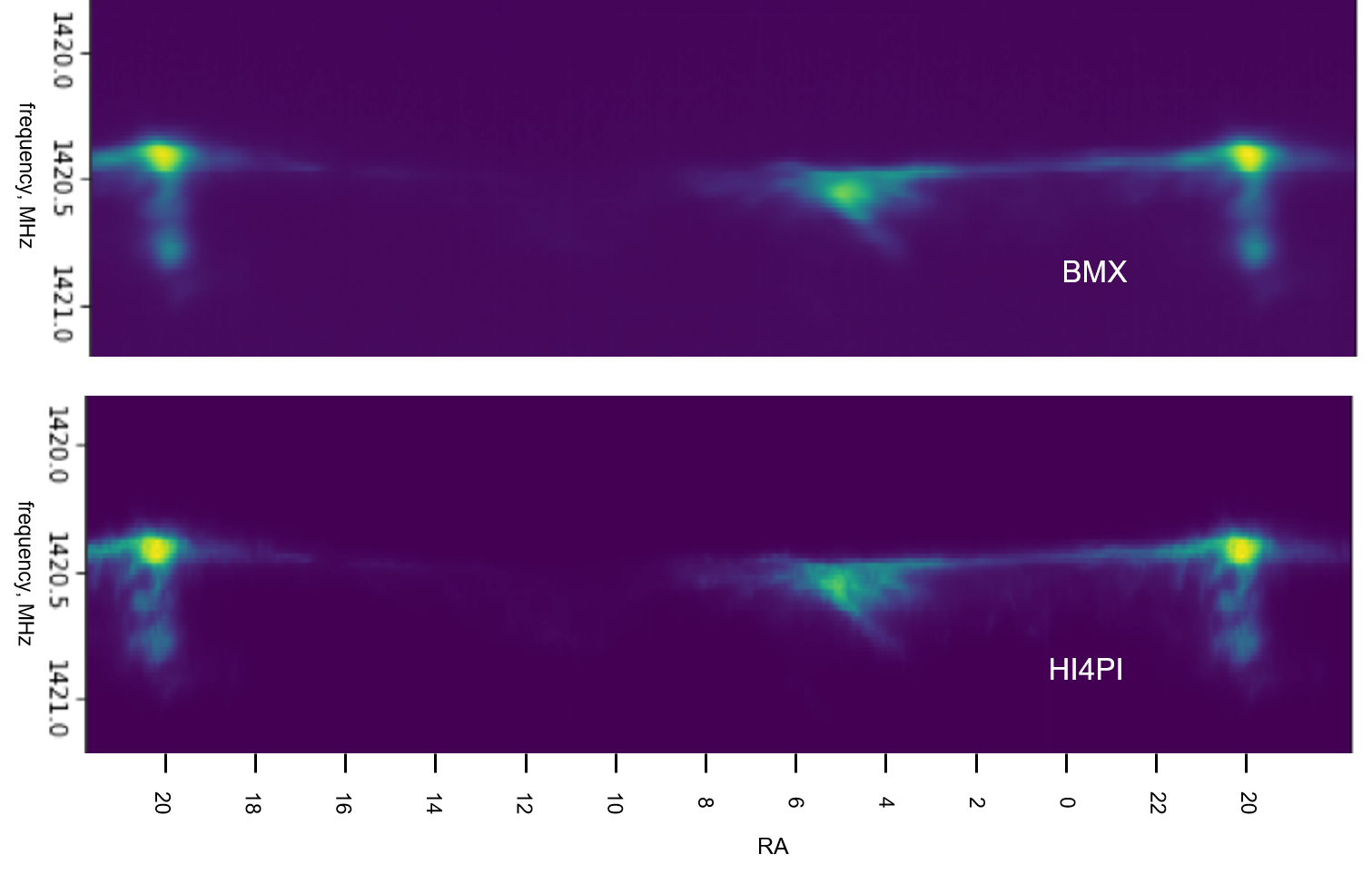}
   \end{tabular}
   \end{center}
   \caption[] 
   { \label{fig:MW1}
  Milky Way map in RA and frequency for the BMX strip. Top map is BMX data, bottom is rebinned data from the HI4PI survey. Note, both maps start at 18h RA (left). After 24 hours, the BMX map extends to 22h the following day while the HI4PI map simply wraps around. A more detailed structure comparison is given in Fig.\ref{fig:MW2}}.
      \end{figure}

\begin{figure} [H]
   \begin{center}
   \begin{tabular}{c} 
   \includegraphics[height=6cm]{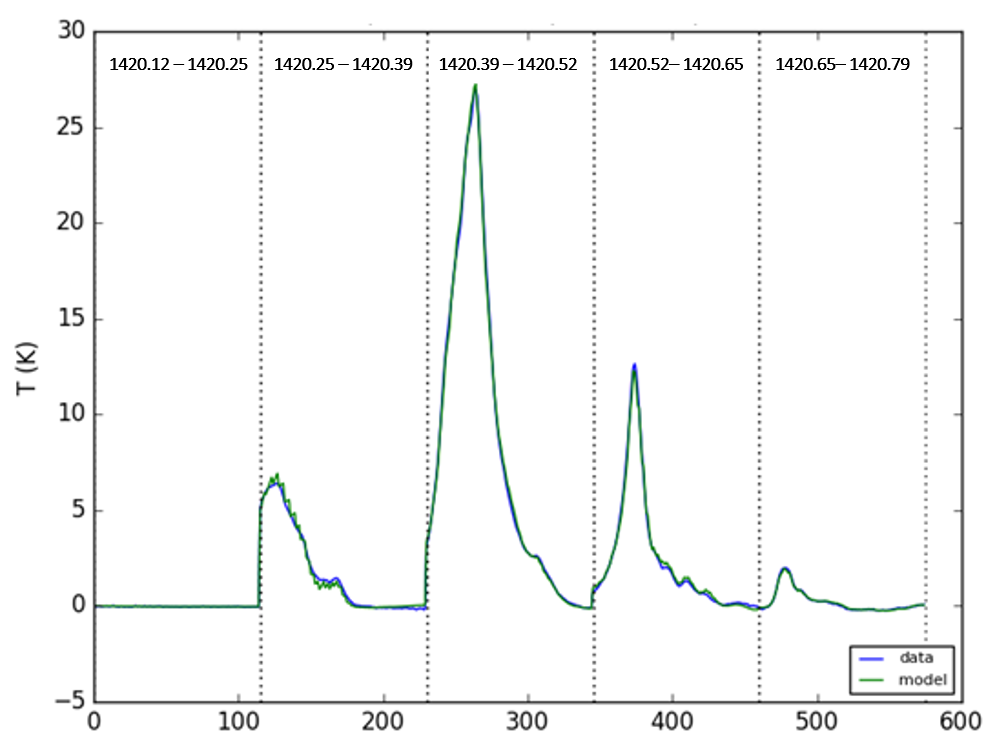}
   \end{tabular}
   \end{center}
   \caption[] 
   { \label{fig:MW2}
   Detail of galactic HI structure. Each of the five sub-panels shows BMX spectrometer data in adjacent 130kHz frequency bins during the MW transit of 1 Dec 2017. The x-axis in each sub-panel covers the same $\sim100$ degrees of RA. After transforming frequencies to a Local Standard of Rest (LSR) which takes into account motions of the Earth and Sun, we fit the data to a Gaussian beam model (shape, pointing, frequency offset) with the HI signal predicted by the HI4PI map. Excellent agreement between the measured (blue) and fitted (green) data is observed for a beam profile with 4 degree FWHM,  small ($\pm 0.5$ deg) departures from nominal zenith pointing, and 20kHz constant frequency offset. }
      \end{figure}

The distribution of HI in the Milky Way galaxy is readily visible in BMX observations using the fine spectral resolution channels centered around the galactic 21cm line. In Fig. \ref{fig:MW1} we show two views of the galactic HI in the BMX strip. The view on the top is BMX data acquired over $\sim 28$ hours starting at 21:45 UTC on 20 October 2020 (data is from the N-Y polarization channel). The western arm of the Milky Way, with nearby Cygnus A, is seen at 20:00 RA near the beginning and end of the run, with the eastern arm visible at 05:00 RA. On the lower panel is the map produced from the high-resolution HI4PI survey \cite{bekhti2016hi4pi}, binned  to  the same RA and frequency resolution as BMX.
A second comparison in shown in Fig. \ref{fig:MW2}, which reveals more detail during passage of the galactic plane around 20:10 UTC on 1 Dec. 2017. For each of five 130-kHz frequency bins, the BMX signal vs time or RA (blue) is fitted to the predicted signal from the corresponding binned RA-frequency sections of the HI4PI map (green). Each section of Fig. \ref{fig:MW2} corresponds to the same $\sim 100$ degrees of RA.  The BMX frequencies were first adjusted to the local system of reference used by HI4PI, which takes into account the motion of the Sun and Earth relative to nearby stars. A beam model with shape (pointing and width), frequency offset, gain, and amplitude offset terms was fit to the data using all frequency bins. Very good fits were obtained for FWHM = 4 degrees, small pointing offsets in RA and Dec  ($\pm 0.5$ deg from zenith), 20kHz frequency offset, and very small offset and gain drifts.

\subsection{GNSS transits}
\begin{figure} [H]
   \begin{center}
   \begin{tabular}{c} 
   \includegraphics[width=\textwidth]{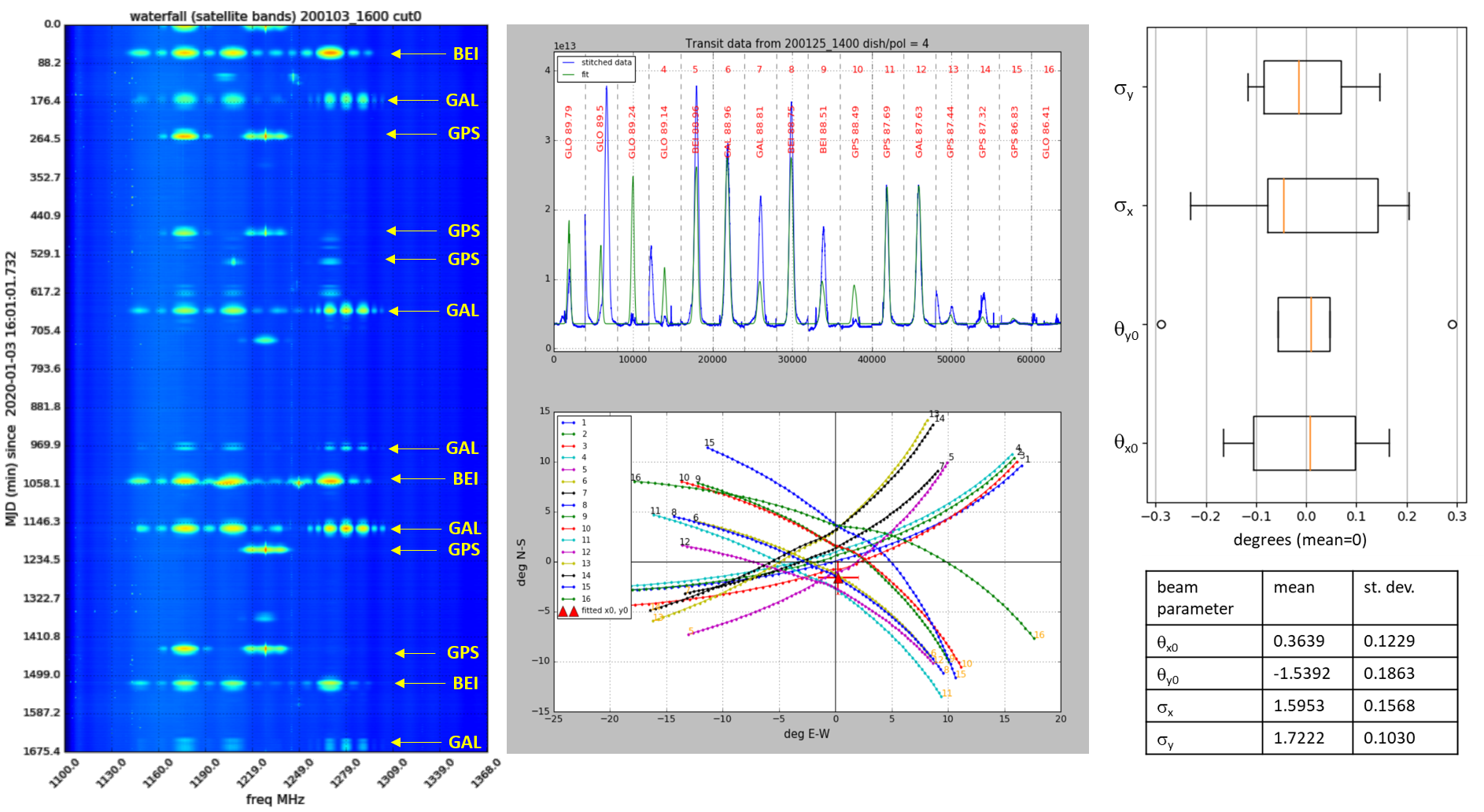}
   \end{tabular}
   \end{center}
   \caption[] 
   { \label{fig:GNSS_fits} 
Left, waterfall data showing signals from GPS, GALILEO, and BEIDOU satellites on 3 Jan 2020. Center top, observed BMX signals (integrated autospectra, blue) for one-hour periods around the predicted transit of 16 satellites on 25 Jan, with fits to simple Gaussian beam model (green). Center bottom, predicted satellite tracks on zenith-centered coordinates with fitted beam center and widths. Right top, statistics of fitted beam parameters over 21-25 Jan 2020, mean-subtracted. Right bottom, resulting parameters (in degrees). Data for West dish, Y polarization.}
   \end{figure}

GNSS satellite transmissions in the L5 (1176.43 MHz) and L2 (1227.6 MHz) bands are observed clearly (Fig. \ref{fig:GNSS_fits}, left) and can be used to measure beam pointing and beam widths. For each dish and polarization, we used the following method:

\begin{itemize}
    \item For each day of observing, identify GNSS satellites that pass near the BMX zenith. We use the two-line element orbit parameters on that day for each satellite to determine local altitude and azimuth as a function of MJD, then sort by maximum altitude and select the 16 satellite transits closest to the BMX nominal (zenith) pointing.
    \item Convert (altitude, azimuth) to a coordinate system centered at zenith with (x,y) axes oriented along (E-W, N-S) directions
    \item Identify windows of length 65  minutes centered around the MJD of transit for each satellite.
    \item Extract the auto-spectrum signal for each window where a satellite is predicted to pass. Integrate over the band of frequencies 1105 - 1369 MHz containing the GNSS transmissions.
    \item Assemble a data cube by stitching together the predicted locations and BMX signals for each of the 16 time windows
        \item Perform a joint fit to all 16 transits using a simple Gaussian beam model of the form \[B(x,y) = Ae^{-((x-x_0)^2/{2\sigma_x^2} + (y-y_0)^2)/2{\sigma_y^2})} + C\]
    Note that the amplitude and offset parameters $A$ and $C$ are treated as invariant and included in the joint fit, even though they may be expected to vary from transit to transit due to satellite transmit power differences and gain drifts. The fit for the shape parameters $x_0, y_0, \sigma_x, \sigma_y$ was found to be insensitive to these variations. Also, frequency dependence is disregarded and the fitted shape parameters may be taken as an average over the 269 MHz GNSS L2+L5 band.
\end{itemize}
An example of the resulting fit for 16 transits occurring on 25 Jan 2020 is shown in the center upper panel of Fig. \ref{fig:GNSS_fits}, along with the predicted tracks for each (center lower panel).  Satellites from GPS, GALILEO, GLONASS, and BEIDOU constellations were observed. Note, not all GPS satellites transmit in the L5 band. Fits were poor for the GLONASS constellation satellites, posibly due to ephemeris errors, but generally good for GPS, GALILEO, and BEIDOU. The extracted shape parameters for the West dish, Y polarization channel over the 5-day period from 21 to 25 January 2020 were found to be stable to better than 0.2$^{\circ}$ (Fig. \ref{fig:GNSS_fits}, right panel).

\subsection{Cygnus-A transits}
\begin{figure} [H]
   \begin{center}
   \begin{tabular}{c} 
   \includegraphics[width=0.65\textwidth]{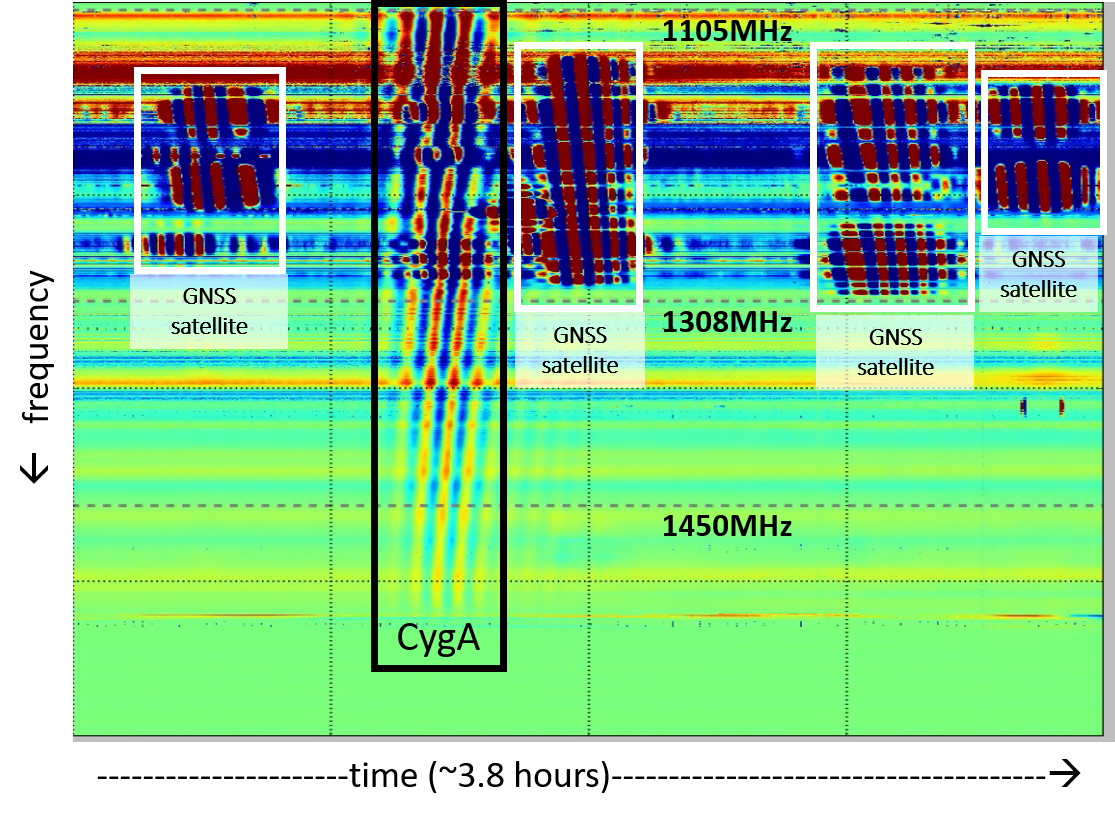}
   \end{tabular}
   \end{center}
   \caption[] 
   { \label{fig:CygA_waterfall} 
Waterfall chart showing cross-correlation amplitude for the East-West dish pair $\sim$3.8 hours of observation near CygA transit. Four GNSS satellites also transited the BMX beam during this period. Dashed horizontal lines mark the frequencies for which CygA fringing was analyzed.}
   \end{figure}

Of the 24 cross-correlation products that a 4-dish, dual-polarization configuration like BMX can produce, we compute only the 12 like-polarization visibilities. Fig.\ref{fig:CygA_waterfall} shows an example waterfall plot of the real part of the East-West, Y-polarization visibilities over a 3.8-hour period near a transit of the radio source Cygnus-A. Dashed horizontal lines are labeled with frequencies for which visibilities were analyzed. Transmissions from four GNSS satellites are seen in the figure in addition to CygA; their signals are confined to the L2 and L5 bands between 1130-1230 MHz while CygA exhibits a falling power-law spectrum. The observed fringes from CygA can be used to estimate beam parameters using fitting methods similar to those used for UAV, GNSS satellite, and galactic HI calibrations, with the advantage that CygA is effectively a point source, its horizon coordinates can be known with high accuracy, and its fringe rate is compatible with BMX's baselines, beam width, frequency bandpass, and integration time.

\begin{figure}[H]
\begin{center}
\includegraphics[width=0.45\textwidth]{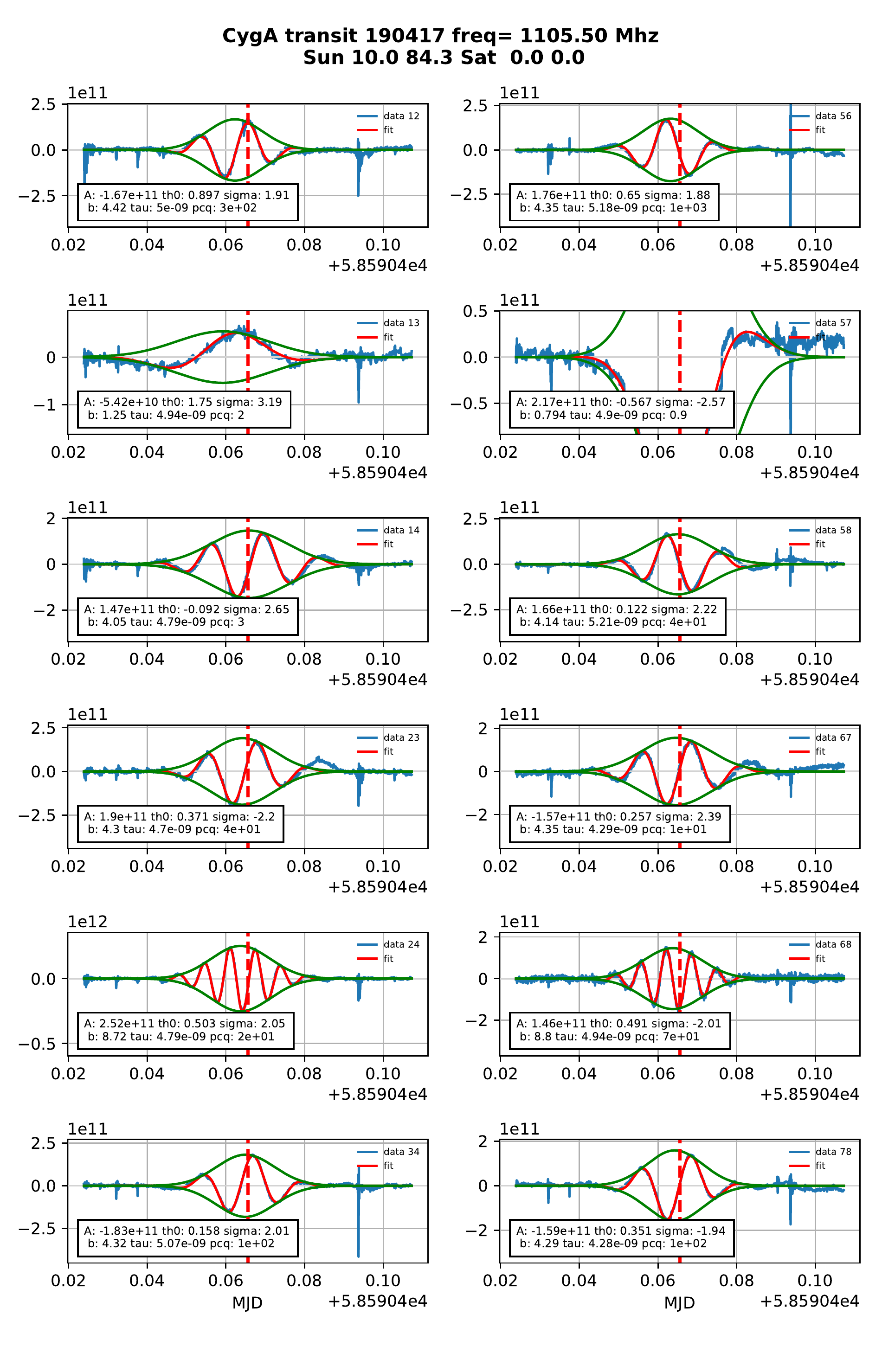}
\includegraphics[width=0.45\textwidth]{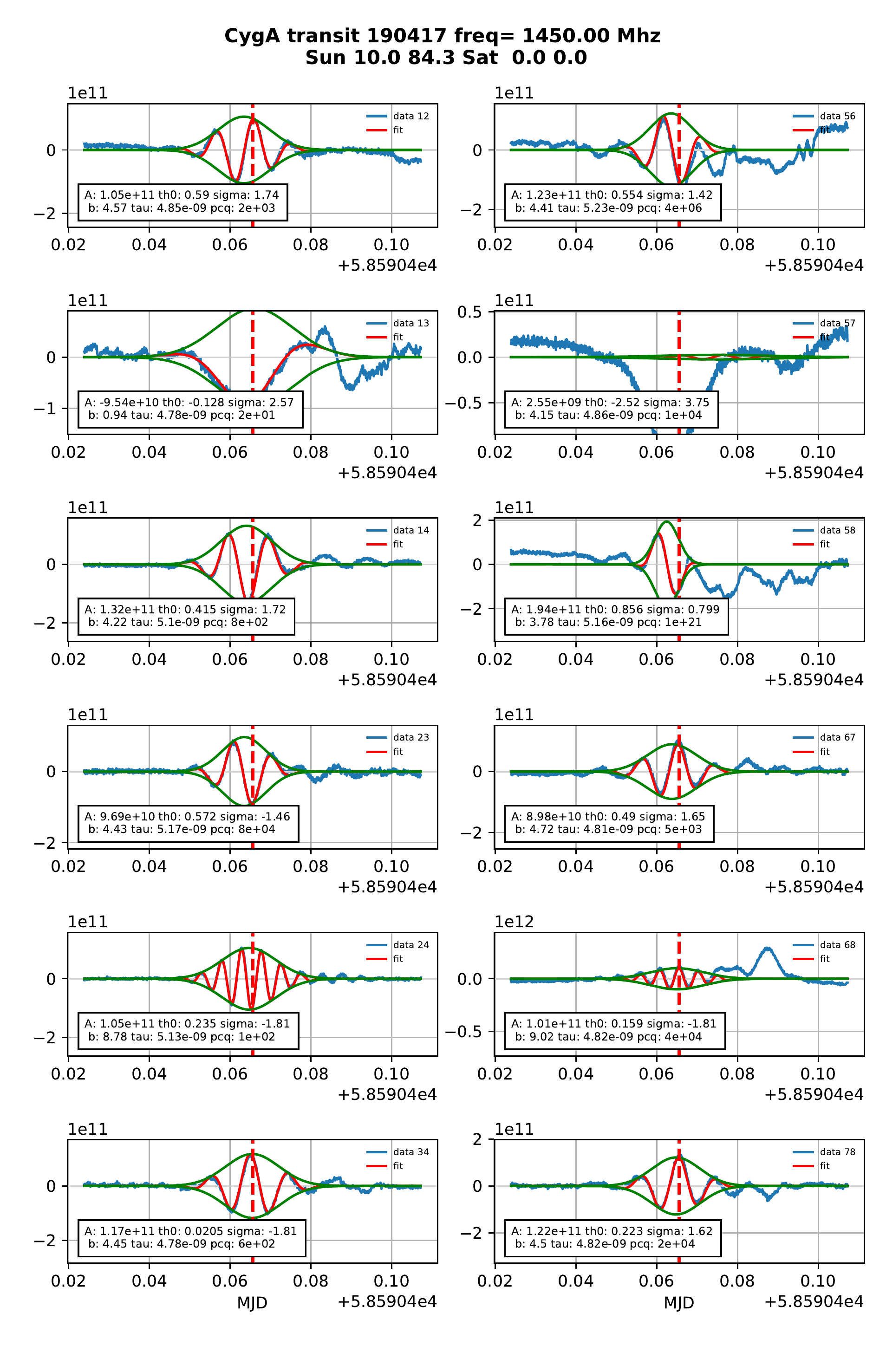}
\end{center}
\caption[]{Visibilities (real part) for a CygA transit on 17 April 2019. Left: at 1105.5 MHz. Right: 1450 MHz. Arrangement of each 6 x 2 panel has like-polarization visibilites in columns and like dish pairs in rows, in the order (N-E), (N-S), (N-W), (E-S), (E-W), (S-W) dish pairs. X-axis is time in MJD. Measured data is in blue; fit in red; envelope of visibility in green. The dashed vertical line is the predicted time of CygA transit over BMX. Inset boxes have the fitted beam parameters: A (amplitude in ADU**2), th0 (RA offset in degrees), b (E-W projection of baseline length in meters), sigma (beam width in degrees), tau (time offset in seconds), and pcq (goodness-of-fit measure). N-S baselines (second row) display very slow fringes, indicative of a small misalignment in the array layout. Overall features show the expected behavior: fringe rate increases and fitted beam widths decrease with frequency; E-W projections for the four diagonally-oriented baselines (rows 1, 3, 4, and 6) are close to one-half the length of the E-W baseline. 
\label{fig:CygA_fringes_2freqs} 
}
\end{figure}

\noindent
Fig. \ref{fig:CygA_fringes_2freqs} shows visibilities for a CygA transit on 17 April 2019, at 1105.5 MHz and 1450 MHz. Each sub-panel shows the 6 like-polarization visibilities for the (N-E), (N-S), (N-W), (E-S), (E-W), and (S-W) dish pairs, along with model fits and fitted parameters. The N-S fringes (second row in Fig.\ref{fig:CygA_fringes_2freqs}) are slow since that baseline is nearly orthogonal to the direction of CygA motion. Otherwise, the visibilities have $SNR >$ 200 and the simple Gaussian model fits well except for those transits where nearby GNSS satellites or RFI spikes perturbed the fit.

\begin{figure}[H]
\begin{center}
\includegraphics[width=0.48\textwidth]{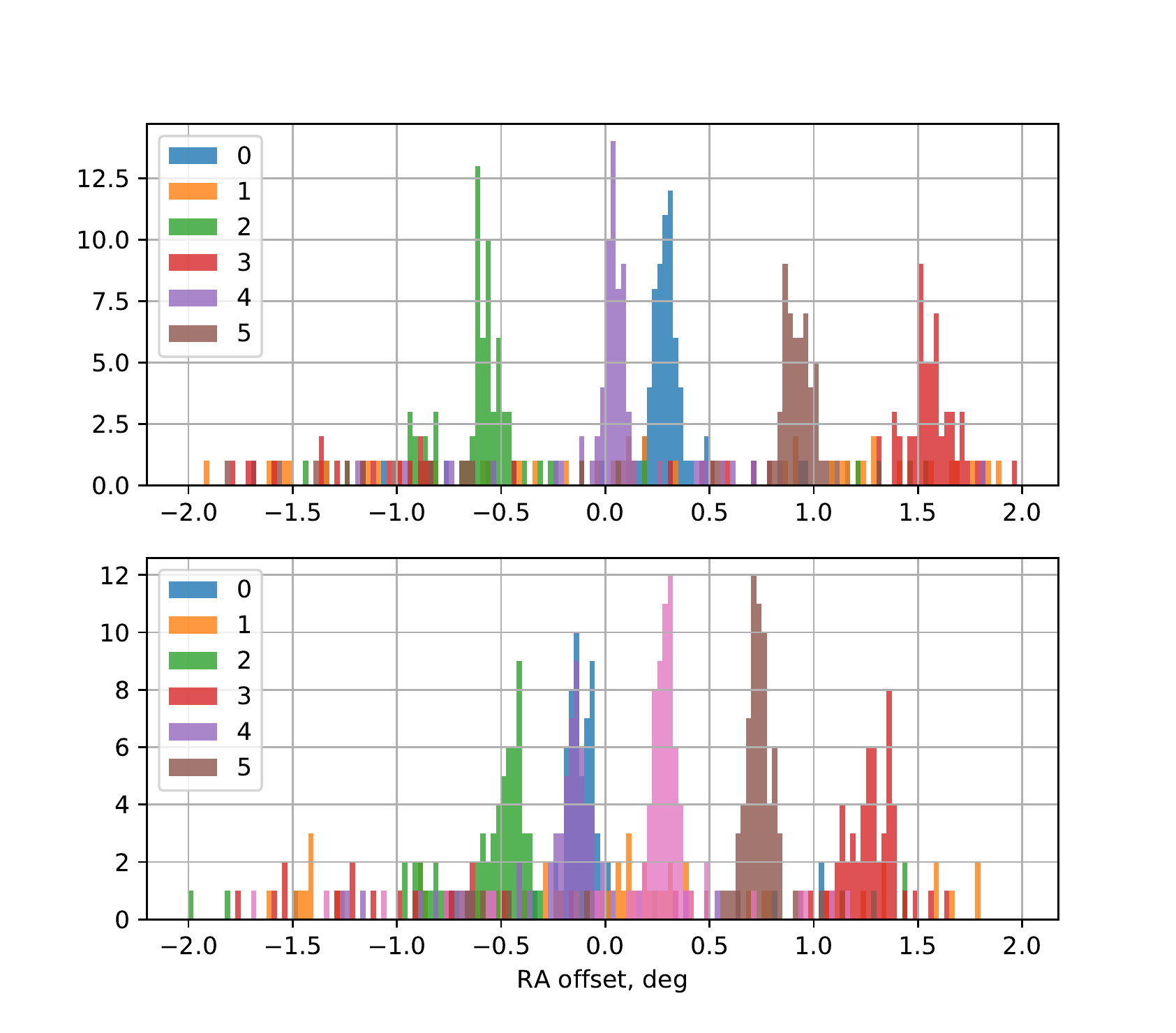}
\includegraphics[width=0.48\textwidth]{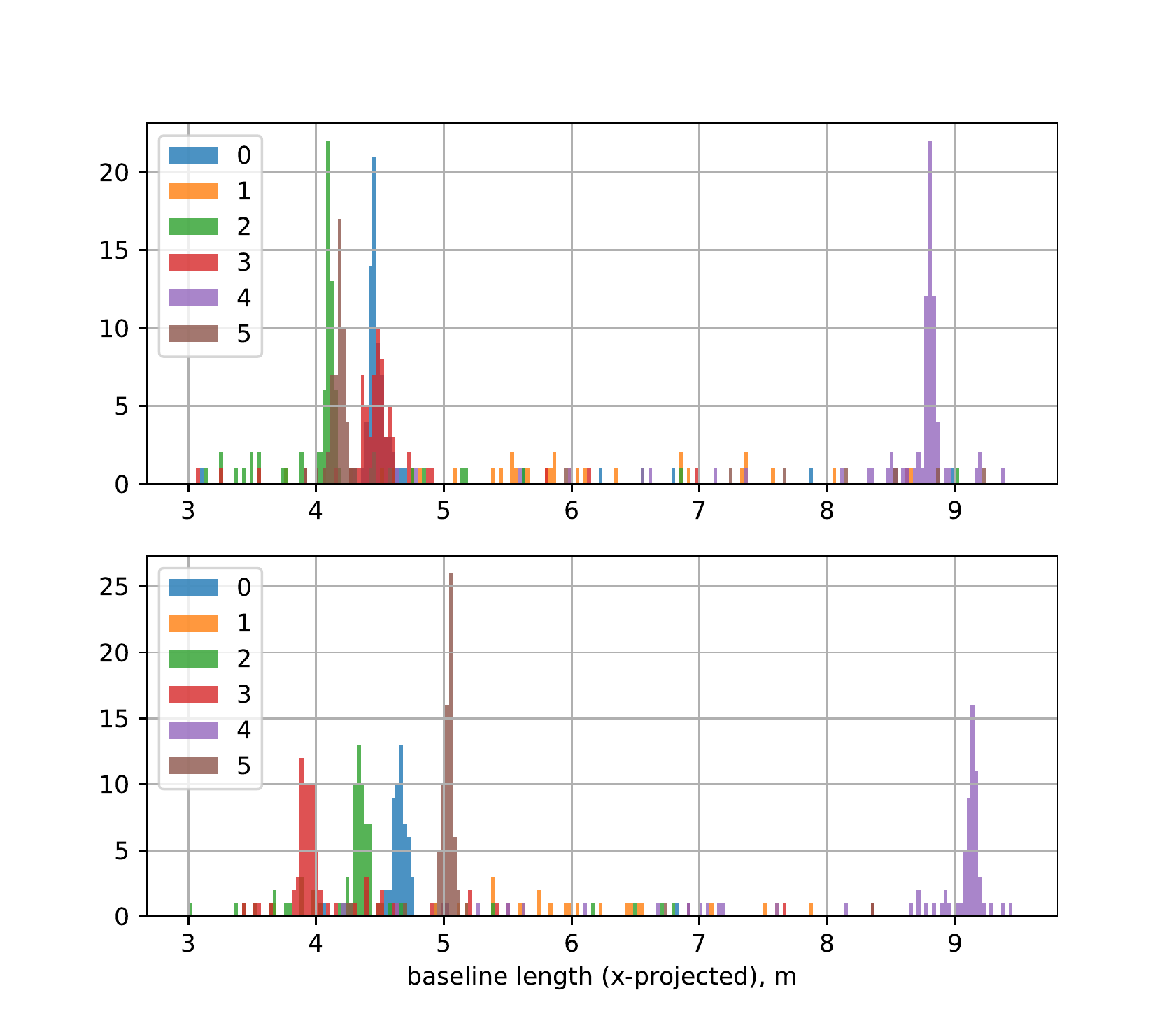}
\end{center}
\caption[]{ Statistics for 81 days of observations in early 2020. Left: Pointing offsets from CygA visibility fitting at  1308 MHz. Upper: X-polarizations. Lower: Y-polarizations. Colors: (blue, orange, green, red, magenta, brown) = (N-E, N-S, N-W, E-S, E-W, S-W) dish pairs.  Right: fitted baselines.
 
\label{fig:CygA_fitHistograms} 
}
\end{figure}

\begin{figure}[H]
\begin{center}
\includegraphics[height=2cm]{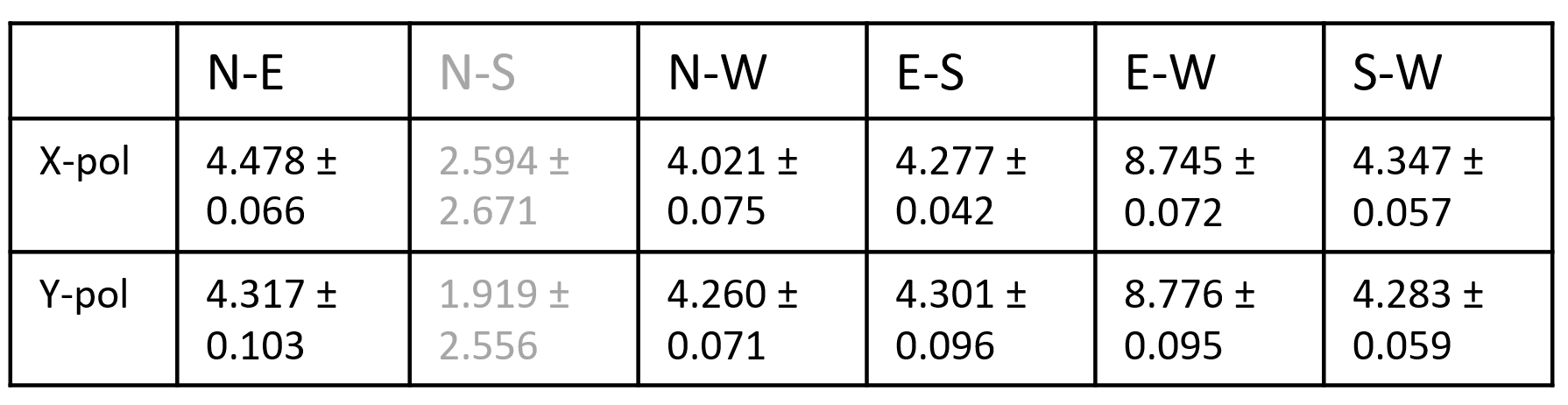}
\end{center}
\caption[]{Statistics of baseline fits to 81 days of CygA transits at 1308 MHz (in meters). Most baselines are stable to better than 2\%. Due to the arrangement of the array, the sum of the N-E and N-W baselines should equal the sum of the E-S and S-W baselines, and both sums should equal the length of the EW baseline. The agreement is within the uncertainty. Note, N-S baseline is poorly fit since it is nearly orthogonal to direction of source motion.
\label{fig:CygA_baseline_stats} 
}
\end{figure}

\noindent
BMX was upgraded from a single-dish spectrometer to a 4-dish interferometer in early 2019, and after commissioning the instrument regular observations began in early 2020. Between 14 January and 17 April we logged 81 days of observations; 16 non-observing days involved weather events, hardware and software maintenance, and coronavirus-related site restrictions starting in late March. Stability of the system is illustrated by the histograms of baseline length and RA offset fitted to CygA transits in Fig.\ref{fig:CygA_fitHistograms}. Fig.\ref{fig:CygA_baseline_stats} gives the statistics for the fitted baselines over this period. After rejecting outliers, pointing and beamwidth fits for a given visibility were generally stable to  0.1 -- 0.2 degrees and baseline fits to within 6 -- 10 cm during this period.
   
\subsection{Calibration Drone Testing}

Beam calibration measurements of BMX were performed in March of 2020 using a drone based radio calibration source developed at Yale University. The noise source was a commercial broadband source from the Ziyang Junying Technology Co. (junying-mw.com), with a 50~kHz $\text{-}$ 1.5~GHz band, which was filtered down to 1.15~GHz $\text{-}$ 1.5~GHz. The source had ENR 90~dB, though it was attenuated down by 43~dB when mapping the main lobe, and 30~dB when mapping the sidelobes, in order to avoid compressing the BMX amplifiers. The drone platform was a Matrice 600 Pro hexacopter produced by DJI. The DJI Realtime Kinematic (RTK) GPS add-on for the Matrice was used to improve the accuracy of the drone positioning. We have verified that the location of the drone is accurately recorded by the RTK system to $<1\mathrm{cm}$ RMS. The transmitting antenna was a biconical Bicolog 20300 antenna from Aaronia. The transmitting antenna was chosen for its uniform gain across the BMX band, and for its isotropic radiation pattern.

Beam calibration was performed by flying in a grid pattern at constant altitude in the farfield of the telescope, with passes along north-south and east-west directions, while holding the heading of the drone constant. Holding the heading constant ensured the polarization angle of the transmitting antenna was maintained as a constant throughout the flight. Two square grids were flown, one 120~m on a side with 6~m spacing between passes to map the main lobe and sidelobes of the beam, and one 70~m on a side with 3.5~m spacing between passes to produce a finer resolution map of the main lobe of the beam. 

\begin{figure} [ht]
   \begin{center}
   \begin{tabular}{c} 
   \includegraphics[width=\textwidth]{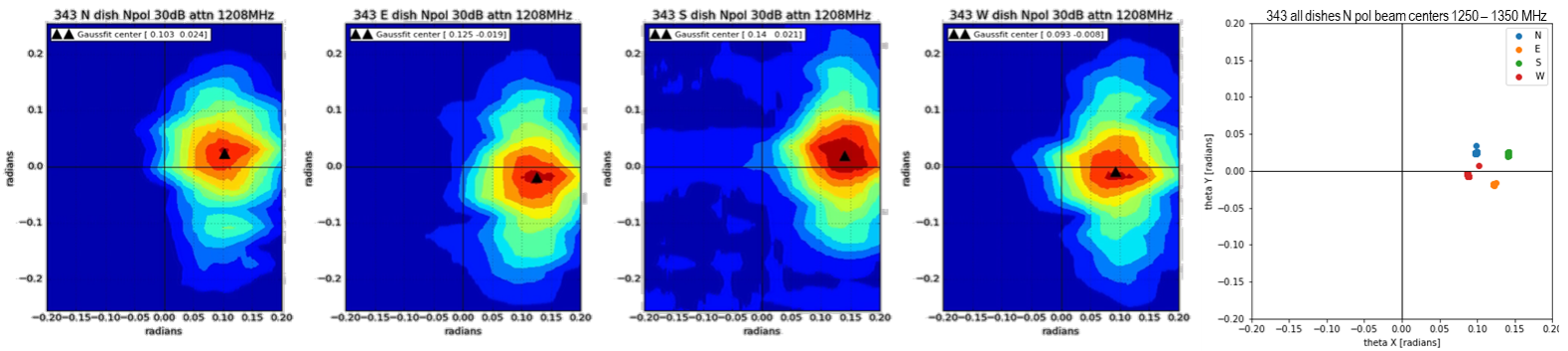}
   \end{tabular}
   \end{center}
   \caption[] 
   { \label{fig:beam_profiles} 
Left, beam profiles for the four dishes at 1208 MHz, interpolated signal for flight 343, a N-S raster at high power with antenna polarization parallel to flight direction. Black triangles are pointing centers from Gaussian fit. Right: pointing centers from Gaussian fits from 1250 -- 1350 MHz.}
   \end{figure}

For each dish, the drone's 3D position in geodetic coordinates was transformed into a 2D coordinate system centered on the zenith. The BMX signal as a function of time was matched to the drone's instantaneous position in that coordinate system. The interpolated beam power patterns are shown in Fig.\ref{fig:beam_profiles}. Each beam was further modeled as either a 2D Gaussian or a symmetric Airy disk. Fits were made for pointing centers and RMS width in the N-S and E-W directions, as functions of frequency. It was found that the absolute time synchronization between the BMX digitizers and the drone RTK clocks were offset at the $<1$s level, by different amounts  for each flight; these were corrected by iterating the fits over a range of time offsets and choosing the offset that resulted in the smallest residuals. All data shown is with time offsets corrected. The right panel of Fig.\ref{fig:beam_profiles} shows the fitted beam pointing centers in 2 MHz steps from 1230 - 1350 MHz. The pointing centers do change with frequency by less than .003 radians, too small to be seen at the scale of the figure. 

\begin{figure} [ht]
   \begin{center}
   \begin{tabular}{c} 
   \includegraphics[width=0.5\textwidth]{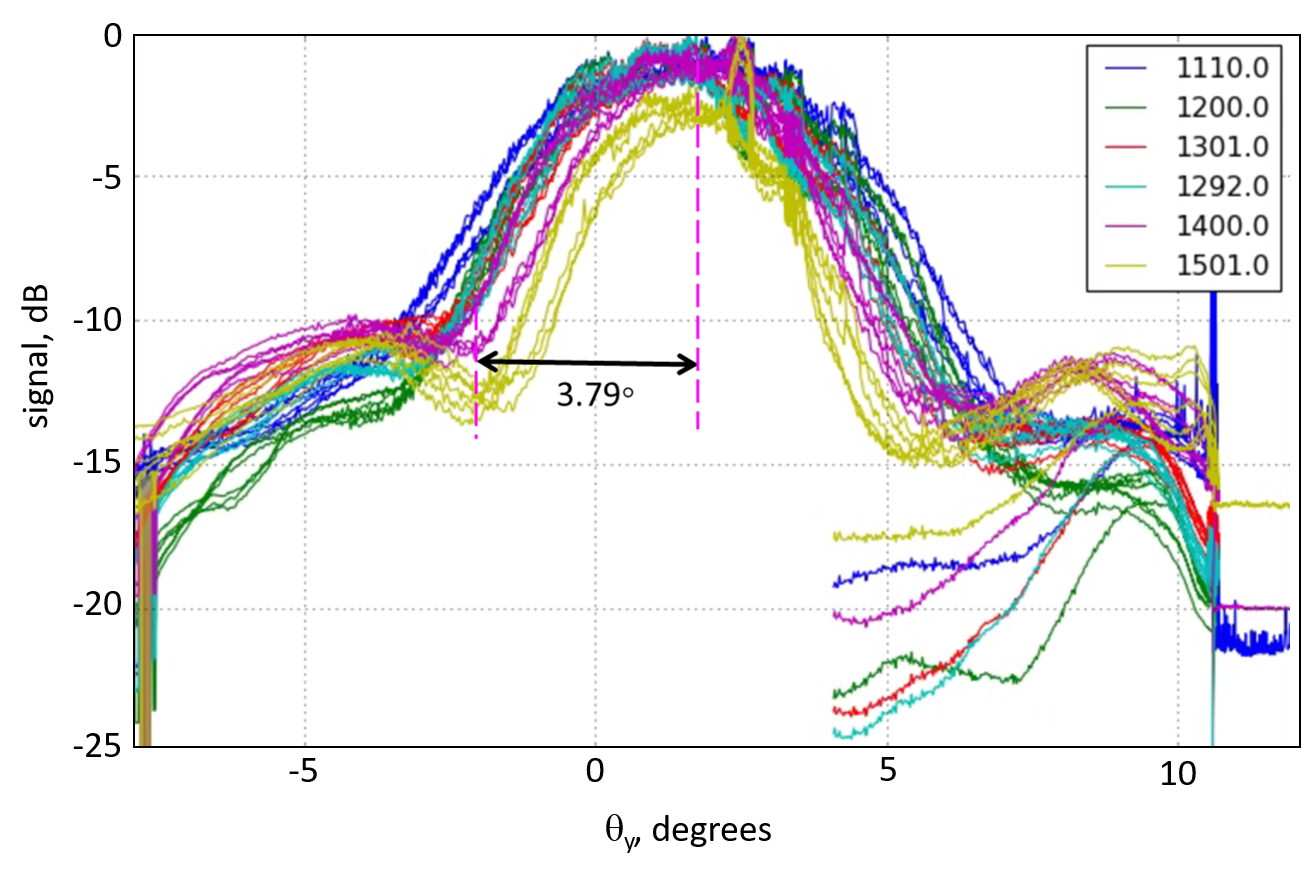}
   \end{tabular}
   \end{center}
   \caption[] 
   { \label{fig:1d_meas_beam} 
Beam profile at 6 frequencies. UAV is flying a wide N-S raster at high transmit power with polarization aligned along flight direction. South dish auto-correlation for the N-S polarization channel is shown. At 1400 MHz, the first sidelobe is visible and the position of the null at 3.79 degrees corresponds to that of the Airy disk from a 3.95m diameter aperture (BMX dishes are nominally 4m-diameter sections of a parent paraboloid).
}
   \end{figure}

Fig.\ref{fig:1d_meas_beam} shows a cut through the observed power along a N-S track near the center of the beam of the S dish. The central lobe narrows with frequency as expected, and the first sidelobe can be seen. The position of the first null at 1400 MHz corresponds to that of an Airy disk from a 3.95m-diameter aperture, very close to the nominal 4m diameter of the BMX dishes.

\section{Summary and Future work}
We have described the BMX interferometer, a testbed for 21cm cosmology research and a pathfinder for the  proposed PUMA intensity mapping array. It incorporates an off-axis optical design with four dual-polarization receivers working in the 1100 - 1550 MHz band, and uses onboard GPU-based correlators to generate eight auto-correlation spectra and 12 cross-correlation visibilities. The system runs in continuous streaming mode and writes about 24GB of data to disk each day. We have presented several applications to demonstrate the performance and stability of the instrument, including beam calibrations using manmade and celestial sources. 

In the near future we plan to investigate a fixed-wing UAV platform for beam calibration, and to develop a compact digitizer and F-engine utilizing the recently developed RFSoC product line from Xilinx. On the analysis side, we have started to develop machine learning techniques for RFI mitigation and will make more detailed use of the GNSS satellites by accessing their modulation codes as well as their RF power for better beam and timing calibration. Finally, we hope to detect the cosmological 21cm signal in cross-correlation with optical galaxy surveys in the redshift range $0 < z < 0.3$.

\section{Acknowledgement}
The authors gratefully acknowledge support from the Laboratory Directed R\&D program at Brookhaven National Laboratory and the Generic Detector R\&D program from the Dept. of Energy Office of High Energy Physics.

\bibliography{main} 
\bibliographystyle{spiebib} 

\end{document}